**Enhancing Knowledge Sharing between Educational Portals**

Maria Teresa Noguera

Fundación Evolución, Argentina

**Corresponding author**: Maria T. Noguera, Fundación Evolución, San Martín 977 4º G, C1004AAS, Ciudad Autónoma de Buenos Aires, Argentina. Email: noguera@fundacionevolucion.org.ar

**Abstract**

Information and knowledge in exchange in public networks is a crucial challenge that needs to be overcome in order to consolidate the benefits associated with such structures. We study the impact of the nature of the information exchanged over the possibilities of success of this process, basing ourselves on the analysis of the information produced by the members of the network of national educational portals RELPE. One of the main challenges that RELPE faces consists in finding effective ways of sharing information that can promote knowledge transfer between members of the network. We argue that a key factor that prevents information sharing is the use of performance metrics by portal responsibles to evaluate the results of their decisions. These metrics are highly sensitive, context-dependent, and produced through non-standardized methods, all of which reduce the willingness of knowledge sharing. We present a different approach: based on the RELPE case, we propose creating a comprehensive information system aimed at providing reliable and timely information in a systematic fashion. We believe that adopting standardized procedures and indicators of less sensitive nature, we can produce information for all partners without the shortcomings of the usual practices.

**Keywords**

**Enhancing Knowledge Sharing between Educational Portals**

*Introduction and Problem Motivation*

The exchange of knowledge in the field of public networks is a crucial challenge they need to overcome in order to consolidate the benefits associated with such structures (Dawes, Cresswell, & Pardo, 2009; Weber & Khademian, 2008). It has been shown that knowledge sharing through social networks helps agencies in the public sector in achieving effective solutions to many of their problems (Dawes, 1996), managing shared resources, creating learning opportunities, and generating innovation (Powell, Koput, & Smith-Doerr, 1996). The process of knowledge sharing is complex and involves cultural and organizational dimensions and aspects related to the nature of the content actually being exchanged (Pardo, Cresswell, Dawes, & Burke, 2004). The nature of the information or knowledge to be exchanged determines the possibilities of success of this process. In order to consolidate partnership in the context of a transnational network of agencies, information sharing requires being of a nature that allows the perception of its value (Zhang, Faerman, & Cresswell, 2006). In this paper we focus on this issue by analyzing the information produced by national educational portals that make up a transnational network of public agencies.

The work presented in this paper is part of the proposal of creating a comprehensive information system for the network of national educational portals RELPE (Red Latinoamericana de Portales Educativos) [1], aimed at providing reliable and timely information in a systematic fashion, through the adoption of standardized procedures. The RELPE network serves as a framework for the discussion that follows, and as a case study that motivates our proposal.

The remainder of this paper is organized as follows. We begin introducing our case study, the RELPE, and the problem that led to its formation as a transnational knowledge network (TKN) (Gharawi & Dawes, 2010). We describe the information produced by the network members in their daily practice, and analyze and discuss its nature and the barriers that such nature imposes to information exchange. Based on the conclusions of this discussion, we then propose a methodology aimed at promoting knowledge exchange in educational portals within a TKN. Our proposal is supported by two groups of metrics, those that produce information about portal management, and metrics that we denote "segmentation metrics", which that allow us to group together the portals according with their activity and relative size. We conclude with a summary of our contribution and an analysis of open research directions in the field. We remark that in this paper we do not address the operational and technical issues related to the production of the metrics proposed [2]. We rather intend to promote a conceptual discussion about the contribution of these metrics as a resource of information that contributes to encourage knowledge sharing between network members.

*The Information Produced by Educational Portals*

The availability of quality digital educational content, together with the provision of telecommunication equipment and infrastructure, are crucial to complete the scheme of public investment in educational technology (Gértrudix Barrio, Carmen Gálvez de la Cuesta, Álvarez García, & del Valle, 2007; Pandian, 2008). The limited development in the production and supply of these contents is a common and well-known problem for many countries (RELPE, 2007b). To address this problem, the network of national educational portals RELPE was created, becoming a community for sharing, storing and free distribution of digital educational content (RELPE, 2010b). The creation of RELPE raised the need to establish a system for gathering information, and producing knowledge to support the development of its members [3].

The RELPE network is characterized by horizontal links between national government agencies involving ministerial levels as well as senior experts in lower-level positions. The Board, President and RELPE Secretariats are held by the ministries of education of each member country, while the implementation of the directives is the responsibility of senior managers and technicians of the member portals (RELPE, 2010b). The network is aimed at solving, by means of collaborative work, a large scale problem as the production of high quality digital educational content (a problem difficult to address in an isolated manner). This requires  the resources developed by some member to be available for all the other network partners, an issue that was technically solved through the definition of standards for cataloguing, and a self-developed content management system (CMS) which allows access to the different portals in the network (RELPE, 2010a).

It has been reported that horizontal networking depends on the exchange of information and knowledge. This exchange can be described as a process that unfolds over time. As agencies begin to work their purposes extend to additional issues (Gharawi & Dawes, 2010). That means, the initial commitment assumed by members of the network opens them the opportunity of sharing other resources, such as information, knowledge and lessons learned during their development. One of the main remaining challenges that managers of transnational network agencies face nowadays, consists in finding effective ways of sharing information, encouraging and promoting knowledge transfer between the members in the network.

Being aware of the problem introduced above, in 2007, we launched a survey and contacted both, managers and technical staff, to determine the practices of production of information about the portal. They were asked about the goals of these practices, the aspects that were measured, the tools and procedures used, and the use of the information produced. This survey involved nine portals, out of the eighteen that composed the network at that time [4].

The analysis of the practices identified allows us to group the information produced in two broad categories. On the one hand, there are those practices that allow obtaining performance information; on the other hand, we have those that produce information related to portal management. We next describe the metrics and indicators in both groups, frequently used by portal responsabiles.

*Portal Performance Information:* Information related to performance was obtained mainly by applying Web Usage Mining techniques (Cooley, 2000) for the analysis of visits, or sequences of page views (click-stream) made by the users [5]. This analysis included aspects such as (a) Activity or traffic; (b) The profiles of registered users; and (c) User valuation of the portal. Among the a*ctivity or traffic* metrics are, for example, the number of pages visited, number of pages views, number of visitors and average visit time. All of these metrics were analyzed for different time periods (e.g. by day of the month, by hour) in order to identify peaks and lows. In the case of sites that had the technology to register their users, metrics were also aimed at knowing the *profiles of registered users*, e.g. age, gender, geographic location, level of education. The portal evolution was studied by means of changes in the number of registered users according to their profile: students, teachers, administrators and researchers. The *user valuation of the portal* was assessed measuring the time spent as a user, frequency of visits, and whenever possible, the participation level indicated by the participation in events promoted by the portal.

We remark that it is a well-known fact that traffic measures have technical limitations, such as the impossibility of knowing the actual number of users visiting the portal: unless a user is registered, visits are counted based on access to the site by the same team and from the same machine. Therefore, the variables above are only approximations to measuring to what extent an educational portal accomplishes one of its goals, which is, reaching as many people as possible.

*Portal Management Information:* Although less widespread than performance metrics, management metrics allow obtaining information about (a) Technology; (b) Entering and exiting behavior of users; (c) The portal's offer of educational topics; (d) Content organization; (e) Positioning of the portal relative to other ones in the network. *Technology* was estimated measuring the average download time of the portal pages, allowing assessing the efficiency of the operation of the portal. Also, usage data was used, whenever possible, to identify the technology of the equipment with which *users* accessed the portal (eg: screen resolution, operating system, browser), in order to adapt the portal design to such technology. *Entering user behavior* was analyzed by identifying the pages in the portal that were used to enter it (usually called "*landing pages*"), and the external sites from where the portal was accessed. *Exiting user behavior* was also studied, and refers to the pages from where the users *exit* the portal (usually called "*exiting pages*"), and the external pages to where the users go after exiting the portal. Regarding content, the *offer of resources* was measured as the number of educational and editorial content per month, per type of user, and per area of curricular knowledge. In general, user preferences was identified analyzing the resources, sites, tools, and services most frequently visited, as well as the text strings entered in the search engine of the portal. In a few cases, the *content organization* (i.e. the organization of the resources offered) was studied, through an indicator of ease of navigation. Finally, the *positioning* of the portal with respect to other ones (inside or outside the network), was assessed measuring the traffic of the site compared to other portals or educational sites [6].

Table 1

*Aspects and Metrics of Information Produced by Portals*

| Information Produced by Portals | | | Indicators and Metrics |
|---|---|---|---|
| Performance | Traffic | | Number of visits, page views, absolute unique visitors, time on site |
| | Users profile | | Number of registered users by age, gender, geographic localization, level of education. |
| | Evolution | | Variations in number of registered users by profiles |
| | User valuation | | Time spent as a user, assiduity, participation, motivation and satisfaction of users of the portal. |
| Management | Technology Used | | Average download time of portal pages |
| | | | Screen resolution, operating systems, browsers |
| | User Behavior | | Input and output behavior of users |
| | Content | Offer | Number and diversity of content |
| | | | Text strings entered in the search engine |
| | | | Resources, sites, tools and services most frequently visited |
| | | Organization | Ease of navigation |
| | Position | | Traffic of the portal compared to other portals or educational sites |

Note that all the measures presented above are associated with practices and site management activities rather than with performance results. That is, operation and organization are related to the management of portal technology; the offer of resources is related to decision about development and/or acquisition of educational content. Finally, comparative information is linked to efforts of positioning, policies and strategies, developed to position the portal on the Web.

*Characterizing the Nature of the Information Produced by Educational Portals*

The nature of the information and knowledge to be shared on a network is a factor of crucial importance for the effectiveness of the process (Zhang et al., 2006). Some of the aspects about this nature, identified in related literature and that will be discussed below are: its sensitivity, its contextualized character, and the lack of standards.

The results of the survey described in the previous section revealed that *information regarding the performance* of the portal gives a rough idea of the capability to attract audience and also of the way in which the products offered by the portal are consumed. Therefore, performance information is used to assess the value of the portals, and to justify the amounts of public resources invested on them. This usage turns performance information into an asset, and, as a consequence, the perception is that it must be considered confidential. That is, sharing this type of information requires high levels of trust. The perceived risks of sharing performance information are related to the loss of control of strategic resources, the possibility of use against the interests of the organization, and the risk of not being corresponded by the sharing partners (Pardo, Gil-Garcia, & Burke, 2006). In summary, taking into account the ***highly sensitive nature*** of performance information, sharing opportunities are limited, at least at an early stage of the process. Trust in social networks requires a high level of social capital to coordinate actions and create common norms and regulations (Coleman, 1988). As in any social network, information sharing does not occur naturally and requires an effort that involves investing in strategies to institutionalize group relationships (Portes, 1998).

On the other hand, although it is well-known that performance information produced by the Web sites is useful for their own administration, its usefulness for sharing in a transnational network of portals is not so clear, as a consequence of its ***highly contextualized nature***. Traffic measures only allow knowing the position of a portal with respect to a "competitor" if the target population is located within the same national context. Traffic metrics such as number of visits or the number of users are affected by local factors such as population size, proportion of the population with access to computers and connectivity, and the level of adoption of these technologies (e.g. percentage of population connected to the Internet). This reduces the perceived value of information for the other portals because each faces different situations.

Opposite to information related to performance, portal management information is ***less sensitive*** and ***possibly more valuable*** for all the network partners due to the fact that it is embedded in the practices of the portal responsibles. The less sensitive nature of information regarding management practices is because it is under the control of the portal reponsibles, helping them in the decision-making process. The knowledge derived from this information would constitute a resource that would increase the repertoire of tools to improve the performance of the portal. Unfortunately this information is produced only for some cases and with procedures that are not systematic, widespread and automated. Moreover, it is produced using techniques that do not allow comparison. Literature on the topic shows that the lack of standard procedures goes against knowledge sharing between agencies (Dawes, 1996). It has been shown that agreed procedures allows solving discrepancies between organizations, increasing partner trust by means of ensuring an even contribution of all network members (Pardo et al., 2004).

The aspects of the information produced by the educational portals mentioned above can become barriers that limit exchanges across the network. These barriers could be overcome if portal managers can perceive that what is proposed to be shared is a contribution that can benefit them in achieving their goals (Dawes et al., 2009). To this end, we argue that it is necessary to rethink the kind of information to be exchanged. In this sense, in the next section we propose a system of metrics oriented to analyze the conditions that lead to the portals to achieve certain results. We will see that these metrics should be applied in the context of a methodology that requires a global vision in which each portal collaborates in the understanding of the conditions that lead to an improvement of the way in which it can achieve its mission. We believe that the process of knowledge exchange involves a negotiation between the actors in order to coordinate what is going to be shared. Thus, our proposal has been designed as a tool that must be able to adapt itself to the needs that may arise from the analysis. Further, the proposal has been conceived as a tool that can be operationalized but must be adapted to the needs which may arise of the analysis and reflective processes that may arise. It is definitely a dynamic tool that must be consolidated together with the consolidation of sharing practices over the network. This proposal is consistent with the vision stating that what people know in an organization and what they can share are deeply embedded in their usual practice (Carlile, 2002).

*A Proposal for Exchangeable Information in the Network of National Educational Portals*

Following the criteria explained in the previous section, we now present a system of metrics that focuses in the production of information concerning portal management, due to its less sensitive nature embedded in the practices of the actors involved. Sharing this knowledge, decisions rationale, experiences, etc, would have a synergic effect in all the network members. This proposal is based on the grounds that the process of building knowledge among organizations involves recursive successive stages which seek to render explicit the tacit knowledge embedded in usual practices, transforming them into knowledge valuable for all the partners (Nonaka, 1994).

As we stated above, in order that this information could be exchanged and its value perceived by others, it has to be produced through standard and systematic procedures. Taking this into account, we identified measures and methodologies aimed at obtaining information for the portal management areas identified in the analyis of uual practices discussed in the previous section. For that, we reviewed the literature on Web Mining (i.e., Web Usage Mining, Web Content Mining, and Link Mining) and, based of this corpus of work, we propose a set of metrics which we denote *Portal Management Metrics* that could be applied systematically and in a standardized way, in order to identify good management practices and the context in which they are carried out. Further, we propose to interpret these metrics with respect to another group of indicators, which we denote *Segmentation Metrics*, that we use to characterize and segment the portals. Both sets of metrics, taken together, allow studying each portal as a case, extracting its experience for exchange and reflection. We next provide a detailed description of both groups of metrics.

Table 2

*Proposal for an Information System for Network of National Educational Portals RELPE*

| Metrics proposed for NNEP Information System | | | |
|---|---|---|---|
| **Management** | Content | Provision | Diversity | Offered |
| | | | | Accessed |
| | | | Richness | Offered |
| | | | | Accessed |
| | | Dynamic | Average age of the contents | |
| | Portal Organization | Depth | | |
| | | Density | | |
| | | Navigability | Portal | |
| | | | Complexity of user navigation | |
| | | Linearity | Portal | |
| | | | of user navigation | |
| | Position | Authoritativeness | | |
| | | Hubness | | |
| | | Bridging | | |
| **Segmentation** | Dynamics of activity | Overall level of demand | | |
| | | Overall level of demand | | |
| | | level of user activity | | |
| | Relative size | Amount of content Portal i / Amount of network content | | |

*Portal Management Metrics*

This group of metrics deals with issues related to three aspects: provision of content, organization of this content within the portal, and positioning and role of the portal in a country's educational Web.

*Provision of Content:* This first aspect includes metrics that allow assessing the portal's resource offer in terms of *provision, and dynamics*. We remark that we consider digital educational content the data -in general multimedia data (text, images, etc.)- stored in Web sites that are available to users (Poblete, 2004), *not including services* such as e-mail or forums. As an example, the portals members of RELPE apply cataloguing standards agreed in advance, to their digital content, based on metadata standards [7]. Each content unit is identified by the metadata "type of educational resource" that describes the resource in the educational context for which it was designed (eg activities, teaching guides, presentations, etc.). Given these data, the provision of digital content can be described using two related standard measures [8]: (a) *Diversity*, which indicates the distribution of content among the different topics that can be found in the portal collection; and (b) *Richness* which indicates the proportion of topics defined by RELPE that are actually present in the content of the portal being studied.

*Diversity* information can be enriched with metrics related to the *diversity of the resources accessed* by users. Since viewing of a resource may be a measure of demand, the temporal variation in the number of unique visitors by content type would describe the audience, identifying and describing seasonal trends. Moreover, information about the *diversity of content offered* and the *diversity of content demanded* allow portal managers to identify the need for adjustments of their content offer. Also, this would allow them to identify whether other network members face similar situations and who could share knowledge on the strategies developed to address them. For instance, the topics identified as having high demand but low offer in a portal can be populated with content developed by other portals for which these portals may have abundant offer and high demand.

Content managed by a portal can be characterized also from the point of view of its *dynamics* using a measure such as the "average age". An analysis across time of this metric may give an idea of the speed at which content is being renewed, thus providing a rough measure of the production efforts of each portal and the degree of up-to-dateness of the resources that the portal contributes to the network. The study of the dynamics of replacement of the contents complements the diversity analysis, allowing identifying the topics or items to which a portal devotes greater efforts. For example, content with high average age and low demand could require replacement; or content with high average age and high demand may indicate that such content is of high quality and it is steadily considered useful. Therefore this content could be analyzed in order to guide in the production process of new content. This is a clear example of information that can be valuable for other members of the network, without affecting sensitive areas.

*Content Organization:* This aspect refers to the order established between the portal pages, based on the URLs and the links that connect them. On this basis, an educational portal can be analyzed as a network that has a certain internal structure with distinctive characteristics (Srivastava, Cooley, Deshpande, & Tan, 2000). This has been studied by the data mining community, and referred as    Web Structure Mining, which is aimed at characterizing the organization of the Web content (Botafogo, Rivlin, & Shneiderman, 1992). Studies of the internal structure of the sites have demonstrated the impact of this structure over the user's probability of effectively finding content within a site, and over the usability of a site (Adhikari & Lemone, 2007; Díaz, 2003; Miller & Remington, 2004). The site structure also influences its possibility of being located and visible to users, and of being indexed by search engines (Baeza Yates, 2004).

To measure and characterize the internal structure of a site we propose four metrics: *depth, density, navigability* and *linearity*. The *depth* of the portal is defined as the average number of links to follow in order to reach a page from the homepage. The *density* of a portal is the number of links on the pages of the portal network, normalized by the total possible links of the site (the result is a value between 0 and 1) (Bordignon & Tolosa, 2006). The shorter the distance between the pages (i.e., the more cohesive a site), the higher the number of links, and therefore the higher the density of the site (Petricek, Escher, Cox, & Margetts, 2006). A value close to 1 means high cohesion of the portal. *Navigability* means that each node of the portal can be easily accessed from other nodes (Pahl, 2001). Unlike cohesion, this measure takes into account the direction of the relationships and the distances from one point to another. Finally, the *linearity* of the portal reveals how well organized it is, and how many choices a user has to take while browsing it (Pahl, 2001). A low value of linearity indicates that the number of choices a user faces is very high, and that the site is unstructured. On the other hand, sites with high linearity could be tedious to navigate. Therefore, medium linearity values are highly desirable.

The combination of these four metrics describes the way in which a portal organizes its content, motivating to study the principles that lead to an effective organization. For example, it is possible to use the counterpart of the last two metrics (navigability and linearity) as measures that indicate the *complexity and nonlinearity of user navigation*. Both seek to describe the behavior of users as a result of the organization of the portal. Both measures have been associated with the understanding that users develop in their interaction with hypermedia resources (McEneaney, 2000). Studies have shown that subjects who are successful in the search processes and in comprehension tests, show high values of complexity and low values of linearity. A high proportion of users with erratic search behavior could indicate the need of adjusting the organization of content. In summary, organizational metrics would provide evidence of aspects that may need adjustment.

*Position Management:* Refers to the management of the hyperlinks and the role that the site accomplishes as a social actor within the educational Web [9]. An educational Web site shares its content with other educational ones in a country through the link structure that connects them. This structure is external to the portal, and it is composed of both, the links incoming to the portal (in-links), and the links outgoing from the portal (out-links).

Links incoming to an educational portal (in-links) are an indication of the value given to the portal contents, and a measure of how the users benefit from such content. Through the link structure, the digital content on the portals becomes a public asset in the scope of a national educational Web, which is an external aspect of the use of educational portals. The number of links outgoing from an educational portal is an indication of the agenda of the portal, i.e., an indication of what is considered important by the portal management staff (Chakrabarti et al., 1999). A portal can play the role of a actor capable of building bridges between Web communities, while influencing the trust, prestige, authority and credibility of other Web sites (Kleinberg, 1998; Park, 2003).

Based on the above, we propose three measures that can be calculated for a collection of educational Web sites in the country to which the portal belongs. The first measure is the level of *authoritativeness*, popularity and prestige of the portal, defined as the incoming degree (Id) or number of incoming links from the other sites being considered. The second measure, *hubness*, measures the capability of a portal to guide and lead their audience to other sites (Chakrabarti et al., 1999). It is measured by the number of outgoing links, and denoted *outgoing degree* (Od).

The third measure proposed, *bridging*, is a combination of the degree (i.e., the sum of the number of incoming and outgoing links), and the number of communities adjacent to the portal, and allows to determine the role of the portal with respect to the educational Web of reference (Scripps, Tan & Esfahanian, 2007). A portal with low degree and high number of adjacent communities is what in the literature of social network analysis (SNA) is called a "bridge" (Burt, 2000, 2001).

A metric that can describe to what extent a site is an *authority* in an educational Web, is a more reliable indicator of the value of the portal than a metric that relies on Web traffic data. It is an indicator of the judgment and recognition to the content provided, produced by the "expert" actors who work in the same context and are probably seeking to satisfy the same target populations. To what extent a portal is a *Hub* depends on the policy that is carried out to manage it. This implies an action plan that considers explicitly establishing links with other actors, and indicates a concern for knowing what the other ones have to offer to the educational communities.

Different to the previous two characteristics, the role of *bridge* that an actor plays is an structural property of a social network, by which a node in  a network serves  as a "broker", building bridges that link different blocks of players with each other, increasing the probability of non-redundant access to information (Burt, 2000). The more heterogeneous are the building blocks being linked, the greater the possibility of distributing resources that can enrich the network (Lin, 1999). Thus, the educational portal, through its external links, can play an important role in disseminating new ideas and content within the country's educational Web. This role, through which the portal exerts influence over other ones, allows to disseminate  ideas that can be relevant an innovative for the different groups (Burt, 2004), and that constitute a more appropriate measure of the value added by the portal, than measures based on the competition for attracting users. At the same time, they provide an alternative view of the portal that goes beyond its primary role as a repository and provider of quality educational content, thus becoming an instrument for linking and disseminating ideas.

*Segmentation Metrics*

The metrics described in the previous section, although valuable, would produce even more interesting information if they were analyzed in light of the particular characteristics of the educational portals. Therefore we propose another group of indicators that account for such characteristics. We denote this group as *segmentation metrics*. These indicators measure the ***dynamics of activity*** of a portal and the ***relative size*** with respect to the size of the portals in the network.

The ***dynamics of activity of a portal*** is given by the variations in a given period of time of the level of gross demand (that is, without user identification). This is evaluated by means of three metrics: *overall level of demand*, the *recency* of users, and the *level of user activity*. The *overall level of demand* is measured as the number of visits, or sequences of views, received by the portal in a certain period of time, without considering the visits from automated agents (crawlers and spiders). The *recency* is defined as the average time between visits by the same visitor in a certain period of time (again, visits from automatic agents are discarded). Finally, the *level of user activity* is measured as the average number of page views per visit calculated as the total sum of page views with respect to the total number of visits in a certain period of time.

The second aspect that we propose to use to account for the diversity of the portals is the ***relative size or relative abundance*** of a portal content, computed as the *amount of content* (without duplication) in the portal, relative to the total amount of network content. In the RELPE cataloging standard, a "content" of a site is uniquely identified by the meta-name [DC:

identifier]. To compute the *amount of content* we need to sum the number of different unique identifiers.

The combination of the two aspects, *dynamics of activity* and *relative size on the portal*, allows establishing an *a priori* typology, with at least four possible groups: (a) Growing portals with large relative size; (b) Growing portals with low relative size; (c) Stable portals with large relative size; (d) Stable portals with small relative size. These categories can serve as a reference to give context and segment the analysis of management measures outlined above.

To illustrate the joint use of metrics in the two groups, let us consider the following example. Two portals A and B in the RELPE network are of similar characteristics (growing and large according to the *segmentation* metrics above), although they organize their content in different ways, resulting in different indicators for *content organization*, which favor portal A. This may trigger an in-depth analysis aimed at finding out whether or not the causes of such differences were known by the responsibles of portal B. During this analysis, the responsibles for Portal B could work together with their colleagues of portal A, and investigate effective organization strategies that have been implemented by the latter. This knowledge could be used by portal B to improve its content organization. Analogously, portal B may have better indicators in other topics (for example, a more diverse content offer), whose analysis may improve the operation of portal A. Note that this joint work would be more difficult (or even impossible) to carry out with indicators of more sensitive nature.

Both groups of metrics, portal management and segmentation metrics, compose a system of metrics oriented to analyze the conditions that lead to certain results, and which must be applied with a global vision in mind, in which each portal collaborates in the understanding of the conditions that lead to their improvement.

*Conclusion*

In this paper we have analyzed the nature of information actually produced by the members of an educational portal network. This information, mainly based on Web site traffic metrics, is commonly used by the portals to support their usual operation. However, we have shown that it is highly sensitive, and more than often, context-dependant. On the other hand, information about portal management is rarely produced, and obtained without following standard methodologies. All of these aspects limit the possibilities and willingness of information sharing among network partners.

With the purpose of overcoming the barriers that prevent knowledge sharing, we proposed a system of metrics to generate information about the management of educational portals (e.g. content offer and organization), and analyzing the results obtained in a context in which results of portals of similar characteristics (e.g. size) could be compared. We argue that this information, yet valuable for management, is much less sensitive than traffic data, and more likely to be shared.

In summary, we claim that portals have valuable, non-sensitive information that is being lost due to the lack of a methodology that encourages knowledge sharing. Therefore, portal responsibles must be convinced that their knowledge contribution to the network will also benefit them in achieving their goals. We expect that this proposal and the collection of metrics presented can contribute to solve the problem of reliably reflecting the effectiveness of the portal management practices that are being carried out.

Footnotes

[1] http://www.relpe.org/

[2] These issues have been developed in other work (Noguera, 2007).

[3] **Full members**:

| | |
|---|---|
| "educ.ar," Argentina; | "Mineduc", Guatemala; |
| "educabolivia", Bolivia; | "Nicaragua Educa", Nicaragua; |
| "Portal do Professor", Brazil; | "Educa Panamá", Panamá; |
| "educarchile", Chile; | "Arandú Rape," Paraguay; |
| "Colombia aprende", Colombia; | "PerúEduca", Peru; |
| "educatico", Costa Rica; | "Educando," Dominican Republic; |
| "Educarecuador", Ecuador; | "Uruguay Educa", Uruguay; |
| "Mi Portal", El Salvador; | "Sepiensa", México. |

**Contributing members**: "CubaEduca", Cuba; "Ministerio de Educación," Guatemala; "hondurasaprende", Honduras; "Portal Educativo Nacional", Venezuela.

**Member of honor**: "ITE" of Spain.

**Associate members** (unofficial public portals): "CEDUCAR", Central America and Dominican Republic; "ATEI", Latin America; "Indágala", Latin America.

[4] "educabolivia", "educarchile", "Colombia aprende", "Mi Portal", "Mineduc", "Sepiensa", "Nicaragua Educa", "Arandú Rape" and "PerúEduca".

[5] Analysis Tools more frequently used were Web-Trends and Webalizer. In some cases these were supplemented with Google Analytics or similar. Web-Trends and Webalizer, are based on parsing, cleaning and analysis of the hits recorded on the Web server log of the portal. On the other hand, in Google Analytics and other similar tools, the data collection technique is based on client-oriented data. They work incorporating executable code into Web pages that allow transfering data to an external database for further analysis. In spite of the current discussion about the limitations implicit in each method of data collection, both can produce essentially the same kinds of metrics.

[6] An example of this is the information provided by Alexa.com.

[7] See standards for cataloging content on the network agreement (RELPE, 2007a).

[8] These are based on the Shannon Index (C. E. Shannon, 1948).

[9] For example, the survey conducted in the framework of the project that originated this work, allowed identifying a set of educational Web sites valued by the community of educational users of one of the portals of RELPE network.

References

Adhikari, V. K., & Lemone, K. (2007). *Hypertext Structural Analysis of Nepali Educational Institution Web-sites*. Paper presented at the 16th Annual World Wide Web Conference

Baeza Yates, R. (2004). Excavando la Web. *El Profesional de la Información,, 13*(1), 4-10.

Bordignon, F., & Tolosa, G. (2006). Caracterización de Espacios Webs Educativos Sudamericanos. *Enlace Informático, 5*(1).

Botafogo, R. A., Rivlin, E., & Shneiderman, B. (1992). Structural Analysis of Hypertexts: Identifying Hierarchies and Useful Metrics. *ACM TransactIons on Information Systems, 10*(2), 142-180.

Burt, R. S. (2000). The network Structure of Social Capital. In B. M. Staw & R. L. Sutton (Eds.), *Research in Organizational Behavior* (Vol. 22). Amsterdam: Elsevier Science.

Burt, R. S. (2001). Structural Holes versus Network Closure as Social Capital. In N. Lin, K. S. Cook & R. S. Burt (Eds.), *Social Capital: Theory and Research*.

Burt, R. S. (2004). Structural Holes and Good Ideas. *The American Journal of Sociology, 110*(2), 349-399.

C. E. Shannon. (1948). A Mathematical Theory of Communication. *The Bell System Technical Journal, 27*, 379-423.

Carlile, P. R. (2002). A Pragmatic View of Knowledge and Boundaries: Boundary Objects in New Product Development. *Organization Science, 13*(4), 442-455.

Coleman, J. S. (1988). Social Capital in the Creation of Human Capital. *The American Journal of Sociology, 94 Supplement*, S95-S120.

Cooley, R. (2000). The Importance of Understanding Web Site Structure and Content when Performing Web Usage Mining. Available at: http://citeseerx.ist.psu.edu/viewdoc/summary?doi=10.1.1.6.7061

Chakrabarti, S., Dom, B., David Gibson, Jon Kleinberg, Ravi Kumar, Prabhakar Raghavan, et al. (1999). Mining the Link Structure of the World Wide Web.

Dawes, S. S. (1996). Interagency information sharing: Expected benefits, manageable risks. *Journal of Policy Analysis and Management, 15*(3), 377-394.

Dawes, S. S., Cresswell, A. M., & Pardo, T. A. (2009). From "Need to Know" to "Need to Share": Tangled Problems, Information Boundaries, and the Building of Public Sector Knowledge Networks. *Public Administration Review, 69*(3), 392-402.

Díaz, P. (2003). Usability of Hypermedia Educational e-Books. *D-Lib Magazine, 9*(3).

Gértrudix Barrio, M., Carmen Gálvez de la Cuesta, M. d., Álvarez García, S., & del Valle, A. G. (2007). Design and Development of Digital Educational Content. In *Computers and Education* (pp. 67-76): Springer Netherlands.

Gharawi, M. A., & Dawes, S. S. (2010). *Conceptualizing Knowledge and Information Sharing in Transnational Governmental Networks*. Albany, NY: Center for Technology in Government.

Kleinberg, J. (1998). *Authoritative sources in a hyperlinked environment.* Paper presented at the Proceedings of The Ninth Annual ACM-SIAM Symposium on Discrete Algorithms, San Francisco, California.

Lin, N. (1999). Building a Network Theory of Social Capital. *Connections, 22*(1), 28-51.

McEneaney, J. E. (2000). *Navigational Correlates of Comprehension in Hypertext.* Paper presented at the Hypertext 2000, San Antonio, TX.

Miller, C. S., & Remington, R. W. (2004). Modeling Information Navigation : Implications for Information Architecture. *Human-Computer Interaction, 19*(3).

Noguera, M. T. (2007, June). Propuesta Metodólogica para el Componente Automático del Observatorio RELPE. *Informe Técnico Nº4*  Retrieved December, 2010, from http://fundacionevolucion.org.ar/investigacion/uploads/groups/0012_group/doc/Informe%20Tecnico%20Componente%20Automatico.pdf

Nonaka, I. (1994). A Dynamic Theory of Organizational Knowledge Creation. *Organization Science, 5*(1), 14-37.

Pahl, C. (2001). *The Evaluation of Educational Service Integration in Integrated Virtual Courses*. Paper presented at the Proceedings of the 2001 Symposium on Applications and the Internet-Workshops

Pandian, M. P. (2008). Digital Knowledge Resources. *Policy Futures in Education, 6*(1), 22-38.

Pardo, T. A., Cresswell, A. M., Dawes, S. S., & Burke, G. B. (2004). *Modeling the Social and Technical Processes of Interorganizational Information Integration.* Paper presented at the 37th Annual Hawaii International Conference on System Sciences (HICSS'04) - Track 5, Big Island, Hawaii.

Pardo, T. A., Gil-Garcia, J. R., & Burke, G. B. (2006). Building Response Capacity through Cross-boundary Information Sharing: The Critical Role of Trust. In P. Cunningham & M. Cunningham (Eds.), *Exploiting the Knowledge Economy: Issues, Applications, Case Studies*. Amsterdam: IOS Press.

Park, H. W. (2003). Hyperlink Network Analysis: A New Method for the Study of Social Structure on the Web. *Connections, 25*(1), 49-61.

Petricek, V., Escher, T., Cox, I. J., & Margetts, H. (2006). *The Web Structure of E-Government - Developing a Methodology for Quantitative Evaluation*. Paper presented at the International World Wide Web Conference Committee (IW3C2).

Poblete, B. (2004). *Herramienta de Minería de Consultas para el Diseño del Contenido y la Estructura de un Sitio Web.* Universidad Nacional de Chile, Santiago de Chile.

Portes, A. (1998). Social Capital: Its Origins and Applications in Modern Sociology. *Annual Review of Sociology, 24*, 1-24.

Powell, W. W., Koput, K. W., & Smith-Doerr, L. (1996). Interorganizational Collaboration and the Locus of Innovation: Networks of Learning in Biotechnology. *Administrative Science Quarterly, 41*(1), 116-145.

RELPE. (2007a). Normas para la Catalogación de Contenidos Educativos. *Documento Técnico Nº1*  Retrieved November, 2010, from http://www.relpe.org/documentos-tecnicos/documento-tecnico-1/

RELPE. (2007b). Stategic Plan. *Consolidating and Integrating the Education Portals Network and Latin America Schoolnets* Retrieved November, 2010, from http://www.idrc.ca/en/ev-123164-201-1-DO_TOPIC.html

RELPE. (2010a). *RELPE: Experiencias Exitosas de Trabajo Colaborativo*: Red Latinamericana de Portales Educativos.

RELPE. (2010b). RELPE: Red Latinoamericana de Portales Educativos [Electronic Version]. *Serie Portales Educativos Latinoamericanos y El Trabajo Colaborativo*, Retrieved November 2010 from http://www.relpe.org/wp-content/uploads/2010/11/Publicaci%C3%B3n-RELPE-1_FINAL.pdf.

Srivastava, J., Cooley, R., Deshpande, M., & Tan, P. (2000). Web Usage Mining: Discovery and Applications of Usage Patterns from Web Data. *SIGKDD Explorations, 1*(2), 12-23.

Weber, E. P., & Khademian, A. M. (2008). Wicked Problems, Knowledge Challenges, and Collaborative Capacity Builders in Network Settings. *Public Administration Review, 68*(2), 334-349.

Zhang, J., Faerman, S. R., & Cresswell, A. M. (2006, January 4-7). *The Effect of Organizational/ Technological Factors and the Nature of Knowledge on Knowledge Sharing.* Paper presented at the Thirty-Ninth Annual Hawaii International Conference on System Sciences, Kauai, Hawaii.